\documentclass[conference]{IEEEtran}
\IEEEoverridecommandlockouts
\usepackage{cite}
\usepackage{amsmath,amssymb,amsfonts}
\usepackage{algorithmic}
\usepackage{graphicx}
\usepackage{textcomp}
\usepackage{xcolor}
\usepackage{hyperref}
\usepackage{multirow}
\usepackage{booktabs}
\usepackage{algorithm}
\usepackage{algorithmic}
\usepackage{subfigure} 
\def\BibTeX{{\rm B\kern-.05em{\sc i\kern-.025em b}\kern-.08em
    T\kern-.1667em\lower.7ex\hbox{E}\kern-.125emX}}
\begin{document}

\title{DCHT: Deep Complex Hybrid Transformer for Speech Enhancement\\
}

 
 \author{\IEEEauthorblockN{Jialu Li\IEEEauthorrefmark{1}, Junhui Li\IEEEauthorrefmark{2}, Pu Wang\IEEEauthorrefmark{3},  Youshan Zhang \IEEEauthorrefmark{4}} 
\IEEEauthorblockA{\IEEEauthorblockA{\IEEEauthorrefmark{1}School of public policy, Cornell University, Ithaca, NY, USA. Email: jl4284@cornell.edu}
\IEEEauthorrefmark{2}\IEEEauthorrefmark{2}Department of Mathematics, School of Science, University of Science and Technology, Liaoning, Anshan, China\\
Email: \IEEEauthorrefmark{2}120203803006@stu.ustl.edu.cn, \IEEEauthorrefmark{3}Junhui\_lee@foxmail.com }

\IEEEauthorblockA{\IEEEauthorrefmark{4}Department of Artificial Intelligence and Computer Science,  Yeshiva University, New York, NY, USA\\
Email: youshan.zhang@yu.edu}}

\maketitle

\begin{abstract}
Most of the current deep learning-based approaches for speech enhancement only operate in the spectrogram or waveform domain. Although a cross-domain transformer combining waveform- and spectrogram-domain inputs has been proposed, its performance can be further improved. In this paper, we present a novel deep complex hybrid transformer that integrates both spectrogram and waveform domains approaches to improve the performance of speech enhancement. The proposed model consists of two parts: a complex Swin-Unet in the spectrogram domain and a dual-path transformer network (DPTnet) in the waveform domain. We first construct a complex Swin-Unet network in the spectrogram domain and perform speech enhancement in the complex audio spectrum. We then introduce improved DPT by adding memory-compressed attention. Our model is capable of learning multi-domain features to reduce existing noise on different domains in a complementary way. The experimental results on the BirdSoundsDenoising dataset and the VCTK+DEMAND dataset indicate that our method can achieve better performance compared to state-of-the-art methods.
\end{abstract}


\section{Introduction}
Speech enhancement (SE) is a challenging task in the field of voice communication since it aims to recover enhanced speech from speech interference with background noise in life. Many voice communication systems may be influenced by background noise, including automatic speech recognition, vehicles, and multi-party conferencing equipment\cite{lu2022conditional}. In response to this issue, a variety of speech enhancement algorithms have been proposed to somewhat minimize noise interference. To address issues such as performance degradation or the challenge of modeling realistic and complicated circumstances, these speech enhancement methods still need to be improved.

Recently, deep neural networks (DNNs) have demonstrated potential for speech enhancement. DNNs for speech enhancement is a crucial front-end approach that takes noisy speech as an input and creates an enhanced speech output for better speech quality and intelligibility. In general, current methods for speech enhancement can be classified into two main categories based on the type of model input. The first category utilizes raw time-domain waveforms, while the latter relies on time-frequency (TF) domain representations. Time-domain methods generally use an end-to-end model to directly estimate and output a clean waveform by taking audio data in the time domain as raw waveform input~\cite{pascual2017segan}. For TF domain methods, they first apply a short-time Fourier transform (STFT) algorithm to transform the input mixed signal into a complex-valued spectrogram, which is subsequently decomposed into its magnitude and phase components. Conventional TF domain techniques utilize the magnitude as a training target while ignoring the phase. Estimated audio signals are then reconstructed via the inverse short-time Fourier transform (ISTFT). However, the frameworks of the time domain are unable to accurately capture speech phonetics in the frequency domain due to the absence of direct frequency representation~\cite{cao2022cmgan}. Besides, most TF-domain methods only take magnitude as input to real-valued parametric models while ignoring complex-valued phases due to the difficulty in estimation, which may limit performance.

In this paper, we investigate the potential of fully end-to-end speech enhancement systems that combine the fragmented benefits of time-domain and TF-domain audio representations, which specialize in tackling specific types of noise. We design a novel speech enhancement transformer framework that involves two speech signal processing modules: a proposed variant of Swin-Unet, named Deep Complex Swin-Unet (DCSUnet), in the time-frequency domain, and an improved dual-path transformer network (DPTnet) in the time-domain. We also introduce a novel cross-domain loss function combining both time- and time-frequency domain losses. Our contributions are as follows:
\begin{itemize}
    \item We propose a novel complex transformer architecture, deep complex Swin-Unet, which combines the advantages of both complex-valued networks and Swin-Net, achieving state-of-the-art performance. We extend the transformer network operator to the complex domain so that it can efficiently model the correlation between the elements of real and imaginary parts of the complex sequence features for a potentially richer representation.
    \item We improve DPTnet by adding memory-compressed attention and propose a hybrid network that combines complex Swin-Net with improved DPTnet named DCHT. The model has a new loss function and performs better than other state-of-the-art audio denoising algorithms on the Voice Bank+DEMAND dataset and the BirdSoundsDenoising dataset.
    \item  Our model utilizes time- and TF-domain representation as input to exploit multi-scale features and effectively extract both time-domain and TF-domain information. The model considers the noise phase information from DCSUnet and extracts local and global audio information from improved DPTnet to exploit the new research techniques for speech enhancement.
\end{itemize}

\section{Related Work}


\textbf{Time-frequency Domain Speech Enhancement.}
Most speech enhancement techniques operate in the time-frequency domain. The complex-valued phase has mostly been ignored, and the majority of methods only estimate the speech magnitude spectrum with reused noisy phase data~\cite{jansson2017singing}. Recently, applying complex spectrograms with complex-valued neural network blocks has become popular because it offers a rich representation. Deep complex U-Net was proposed by incorporating advanced U-Net structures and complex-valued blocks to deal with complex-valued spectrograms~\cite{choi2019phase}. Particularly, the Deep Complex Convolution Recurrent Network (DCCRN) was designed to simulate complex-valued operations using complex networks~\cite{hu2020dccrn}. Complex-valued transformers have also been designed for speech enhancement. Tan et al.~\cite{tan2023cst} presented a complex transformer module with sparse attention to low-SNR speech enhancement tasks. Zhang and Li~\cite{zhang2023complex} converted the audio denoising problem into an image generation task and proposed a complex image generation SwinTransformer network for audio denoising.

\textbf{Time Domain Speech Enhancement.}
The time-domain speech enhancement approach is a method that operates directly on the time-domain mixture speech waveform to predict the clean speech waveform~\cite{macartney2018improved}. Dario et al.~\cite{rethage2018wavenet} proposed an end-to-end learning method for speech denoising based on Wavenet. Baby and Verhulst~\cite{baby2019sergan} introduced a conditional generative adversarial network with a gradient penalty for improved speech enhancement performance. With the popularity of attention mechanisms, Kong et al.~\cite{kong2022speech} presented CleanUNet for speech enhancement, which utilized an encoder-decoder architecture combined with self-attention blocks to refine its bottleneck representations. Transformer can now process input audio sequences in parallel, much like natural language processing. Yu et al.~\cite{yu2022setransformer} proposed a speech enhancement transformer (SETransformer) that takes advantage of Long Short Term Memory (LSTM) and multi-head attention mechanisms to improve speech quality.

\textbf{Hybrid Domain Speech Enhancement.}
Recently, there has been a trend where speech enhancement models combine time and TF domains. Kim et al.~\cite{kim2021kuielab} proposed a two-steam neural network with both TF domain and time-domain branches to separate music stems. Using a bi-U-Net structure, Alexandre et al. ~\cite{defossez2021hybrid} refined the Demucs architecture and built an end-to-end hybrid source separation model blending time and TF domains to improve the quality of music source separation. Furthermore, Simon et al. ~\cite{rouard2022hybrid} introduced a transformer architecture to model the hybrid transform Demucs network for Music Source Separation (MSS) tasks based on Hybrid Demucs.



\section{Method}
We first introduce hybrid transformer models, followed by describing deep complex Swin-Unet and improved DPTNet, respectively. Before getting into details, we assume that the mixture of a speech signal $y(t)$ is the linear sum of the clean speech signal $x(t)$ and the noise speech signal $\varepsilon(t)$, so the noisy speech $y(t)$ can be expressed as:
\begin{equation}
  y(t)=x(t)+\varepsilon(t),
\end{equation}
 A sequence of a mixture signal and a clean signal is defined as 
$Y=\{y_i\}_{i=1}^{N}$ and $X=\{x_i\}_{i=1}^{N}$, where $N$ is the total number of speech signals. Typically, each of the corresponding time-frequency $(k, f)$ noise reduction operates in the time-frequency domain ~\cite{li2020speech}:
\begin{equation}\label{kf}
  Y_{k,f}=X_{k,f}+\epsilon_{k,f},  
\end{equation}
where $Y_{k,f}, X_{k,f}, \epsilon_{k,f}$ is the STFT representation of the time domain signal $y(t)$, $x(t)$, $\varepsilon(t)$ and $k$, $f$ are the time frame index and frequency bins index. In polar coordinates, ~\eqref{kf} becomes:
\begin{equation}
|Y_{k,f}|e^{i\theta_{Y_{k,f}}}=|X_{k,f}|e^{\theta_{X_{k,f}}}+|N_{k,f}| e^{i\theta_{\epsilon_{k,f}}},
\end{equation}
where $|\cdot|$ denotes the magnitude response and $\theta$ denotes the phase response. The imaginary unit is represented by $i$.

A complex-valued convolutional filter is defined as $W=A+iB$ with real-valued matrices $A$ and $B$, which represent the real and imaginary part of a complex convolution kernel, respectively. At the same time, we can get complex output from the complex convolution operation on complex vector $h=x+iy$ with $W$: $W\times h=(A\times x-B\times y)+i(B\times x+A\times y)$.
\subsection{Hybrid Transformer Model}
\begin{figure*}
 \centering	\includegraphics[width=0.7\textwidth]{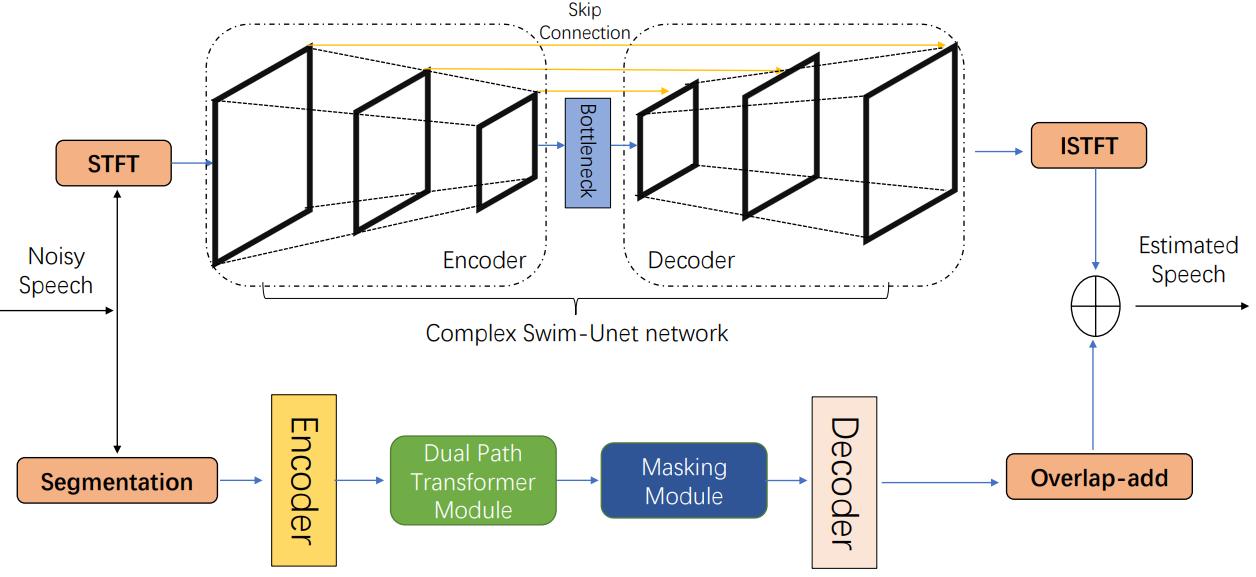}
 	\caption{A overall progress of our proposed hybrid transformer architecture for speech enhancement. }
  \label{Fig:over}
  \vspace{-0.6cm}
\end{figure*}

In this section, we introduce the overall structure of the DCHT model. The overall progress of the proposed model is presented in Figure.~\ref{Fig:over}. The proposed model, which extends transformer network architecture with multi-domain representation, comprises two SE approaches: the complex Swin-Unet and improved DPTnet. The complex Swin-Unet module consists of an encoder, a bottleneck, a decoder, and skip connections. The basic unit of Swin-Unet is the Swin Transformer Block. The proposed complex Swin transformer block is a refined transformer architecture applied in the TF domain. The output of the spectral branch is transformed into a waveform using ISTFT before summed with the output of the temporal branch, giving the actual prediction of the model. The improved DPTnet introduces direct context awareness in the speech sequences modeling by combining memory compression attention, which is efficient for long speech sequence modeling. It learns the order information of the speech sequences without positional encoding by incorporating gated recurrent units (GRU). The 1-D signal from the waveform branch and the 2-D signal from the spectral branch are treated simultaneously. With these modifications, we performed the experiments in Section~\ref{sec:exper}.

\subsection{Complex Swin-Unet Module}
The architecture of the original Swin-Unet is based on the Swin Transformer, which exhibits outstanding performance on computer vision tasks. The modifications made to the original Swin transformer are as follows: The convolutional layers of the Swin transformer blocks are replaced by complex convolutional layers. Complex normalization, complex dropout, and complex linear layers are also implemented to the transformer. To accept complex input and output features, we modified real-value Layernormalization (each token is independently normalized) and real-value softmax into complex Layernormalization and complex softmax. For the activation function, we modified the real-value GeLU into the complex-value GeLU. We use the encoder to extract the hierarchical feature representations from the transformed patch tokens using complex Swin transformer blocks and patch merging layers. The symmetric decoder reshapes each stage feature map and restores the resolution of the feature maps to the input resolution. Skip connections are used to combine multiscale information from the encoder and context features from the decoder to improve the reconstruction of the denoised images.

In practice, the target has to be estimated. Choosing an appropriate training target is crucial for supervised learning since it is directly related to the underlying computational target. The target magnitude spectrum of clean speech is a commonly used training target in typical mapping-based approaches. We consider the complex short-time spectrogram of the noisy speech $Y$, the noise $N$, and the clean speech $X$, obtained via the discrete Fourier transform of windowed frames of the raw signals. The estimated speech spectrogram $\hat{Y}_{k,f}$ is computed by multiplying the estimated mask $\hat{M}_{k,f}$ to the input spectrogram $X_{k,f}$ as Eq.~\eqref{mkf}. More formally, we need to train a prediction model $F$ to predict the mask, and the output is $\hat{M}_{k,f}=F(X_{k,f})$. When the mask connection is not applied, the network directly outputs the estimated complex spectrum, i.e., $\hat{Y}_{k,f}=F_{k,f}$.
\begin{equation}\label{mkf}
\hat{Y}_{k,f}=\hat{M}_{k,f}\cdot X_{k,f}=|\hat{M}_{k,f}| \cdot |X_{k,f}| \cdot e^{i(\theta_{\hat{M}_{k,f}}+\theta_{X_{k,f}})}.
\end{equation}

By applying an additional bounding process, the proposed complex-valued mask $\hat{M}_{k,f}$ is expressed as follows:
\begin{equation}\label{mask}
\begin{aligned}
&\hat{M}_{k,f}=|\hat{M}_{k,f}|\cdot e^{i\theta_{M_{k,f}}}=\hat{M}_{k,f}^{mag} \cdot \hat{M}_{k,f}^{phase},\\
&\hat{M}_{k,f}^{mag}=tanh(|F_{k,f}|),\hat{M}_{k,f}^{phase}=F_{k,f}/|F_{k,f}|.
\end{aligned}
\end{equation}

\begin{figure}
\centering
\includegraphics[width=0.5\textwidth]{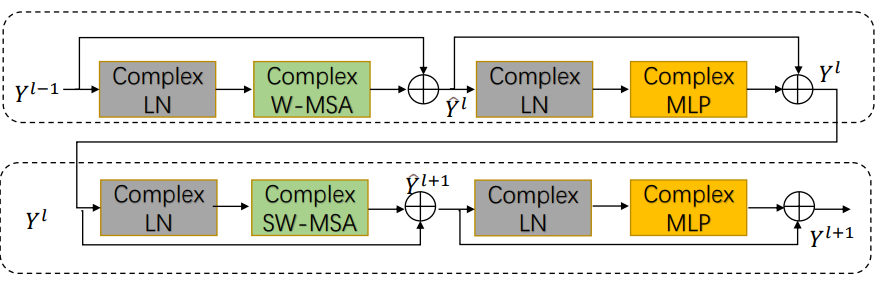}
\caption{The architecture of Complex Swin Transformer block module.}\label{Fig:swinblock}
\end{figure}

\textbf{Complex Swin Transformer Block}
The complex Swin transformer block (CSTB) is responsible for feature representation learning. Unlike the conventional multi-head self-attention-based transformer, the  Swin Transformer~\cite{liu2021swin} has a hierarchical architecture whose representation is computed using shifted-windows multi-head self-attention. In Figure.~\ref{Fig:swinblock}, each CSTB mainly consists of two successive units: the complex window multi-head self-attention (complex W-MSA) unit and the complex shifted-window multi-head self-attention (complex SW-MSA) unit. Each complex LayerNorm (complex LN) is implemented before the complex MSA module, and each complex 2-layer MLP with complex non-linearity GELU and the remaining connections are implemented before and after each module.

On the basis of the shifted window partitioning mechanism, the whole sequential complex Swin transformer blocks are computed as:
\begin{equation}
\begin{aligned}
   \hat{Y}^L & =complex \quad W-MSA (cLN(Y^{L-1}))+Y^{L-1},\\
   Y^{L} &=complex\ MLP(cLN(\hat{Y}^{L}))+\hat{Y}^{L},\\
   \hat{Y}^{L+1} & =complex\ SW-MSA(cLN(Y^{L}))+Y^{L},\\
   Y^{L+1} & =complex\ MLP(cLN(\hat{Y}^{L+1}))+\hat{Y}^{L+1}.
\end{aligned}
\end{equation}

\subsection{Improve Dual-path Transformer module}
In this section, we use the improve Dual-path transformer neural network to train the raw speech waveform input~\cite{wang2021tstnn}. This network is suggested for processing long-distance speech sequences. It is inspired by the transformer's ability to interpret sequences and the dual-path network's ability to gather contextual data~\cite{luo2020dual}. DPTnet consists of an encoder, a DPTnet module (DPTM), a mask module, and a decoder, as shown in Figure~\ref{Fig:over}. DPTM is composed of four stacked dual-path transformer blocks to efficiently extract local and global information. Due to its few trainable parameters, the model complexity of DPTnet is significantly lower.

\begin{figure}
	\centering
         \subfigure[The Improve Dual-path transformer block.]{\label{Fig:DPF}
		\includegraphics[width=0.6\linewidth]{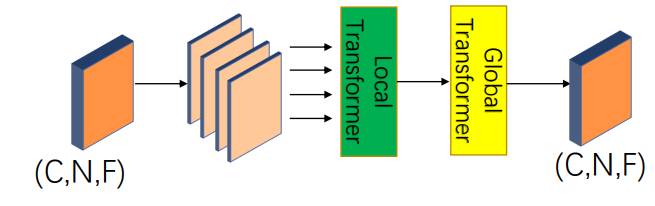}}
	   \subfigure[ Transformer architecture]{\label{Fig:TA}
		\includegraphics[width=0.6\linewidth]{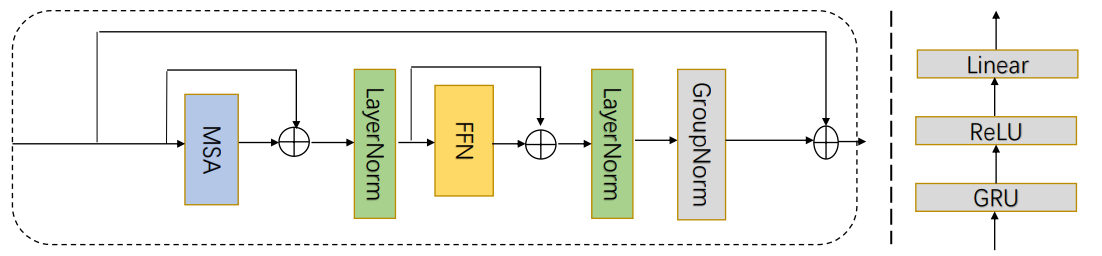}}
	\caption{(a) The Improve Dual-path transformer block module; (b) Transformer architecture.}
\end{figure}

\textbf{Dual-Path Transformer Block}. In the DPTnet extraction module, we substitute the Dual-Path Transformer Block (DPTB) for the traditional transformer block. As shown in Figure~\ref{Fig:DPF}, DPTB is between the encoder and decoder, which has a local transformer and a global transformer to extract local and global contextual information from long-range speech sequences. The input is a 3-D tensor$([C,N,F'])$, and in order to process local information in parallel, the local transformer is first applied to individual chunks. This process works on the input tensor's final dimension $F'$. Subsequently, the global transformer is employed to combine the data from the local transformer's output in order to ascertain global dependency, which is applied to the tensor's dimension $N$. For the DPTB structure, the transformer only uses the encoder part since the input mixtures and output-enhanced sequences have the same length in speech denoising. Since the speech sequence is time-dependent, the transformer structure in DPTB removes the positional encoding part. As shown in Figure~\ref{Fig:TA}, the first fully connected layers of feed-forward and linear normalization are replaced with a GRU layer and layer normalization, respectively. The procedures are defined as follows:
\begin{equation}
    \begin{aligned}
    \hat{z}_i&=LayerNorm(MSA(z_{i-1})+z_{i-1}),\\
    z_i&=LayerNorm(\hat{z}_i+ReLU(GRU(\hat{z}_i)W+b)).
    \end{aligned}
\end{equation}

\subsection{Cross-domain Loss Function}
In this study, our composite model integrates T-F and time streams to extract a set of complementary audio features. Therefore, our loss function consists of time-domain loss and TF-domain loss to fully utilize both feature information. Inspired by the L1 norm of a complex number in ~\cite{braun2021consolidated}, we take the STFT of the audio and use L1 loss over the L1 norm of the STFT coefficients.
The frequency-domain loss is given by:
\begin{equation}
 \begin{aligned}
     loss_{L_1,TF}(y,\hat{y})&=\frac{1}{TF}\sum_{t=0}^{T-1}\sum_{t=0}^{F-1}
     [(|y_r(t,f)|-|\hat{y}_r(t,f)|)\\
     &+(|y_i(t,f)|-|\hat{y}_i(t,f)|)],
 \end{aligned} 
 \end{equation}
where $y$ and $\hat{y}$ denote the spectrum of the clean speech and the spectrum of the enhanced speech, respectively. $r$ and $i$ are the real and imaginary parts of the complex spectrogram. $T$ and $F$ are the number of frames and the number of frequency bins, respectively.

The time-domain loss is based on the energy-conserving loss function proposed, which simultaneously considers clean speech and noise signals. The time-domain loss is defined as follows:
\begin{equation}
    loss_{L_1,T}(y,\hat{y},x,n)=\|y-\hat{y} \|_1+\|n-\hat{n} \|_1,
\end{equation}
where $y$ and $\hat{y}$ are the samples of the clean speech and the enhanced speech, respectively. $\hat{n}=x-\hat{y}$ represents the estimated noise and $\|\cdot \|$ denotes $L_1$ norm.

To properly balance the contribution of these two loss terms and address the scale insensitivity problem, we weigh ($\alpha$) each term proportionally to the energy of each speech. The final form of the loss function is as follows:
\begin{equation}
\begin{aligned}\label{eq:loss}
    loss_{Total}(y,\hat{y})=\alpha loss_{L_1,TF}+(1-\alpha)loss_{L_1,T}.
\end{aligned}
\end{equation}


\section{Experiments}\label{sec:exper}
\subsection{Dataset} 
\textbf{VCTK+DEMAND dataset.} We validate the effectiveness of our proposed model on a small-scale standard speech dataset from~\cite{valentini2016speech}. In this widely used noisy speech database, clean speech datasets are selected from the Voice Bank Corpus~\cite{veaux2013voice}, including a training set of 11,572 utterances from 28 speakers and a test set of 872 utterances from 2 speakers. 

\textbf{BirdSoundsDenoising.} We also train our proposed model on BirdSoundsDenoising. This dataset replaces the usual artificially added noise with natural noises, including wind, waterfalls, rain, etc. In particular, the dataset contains 14,120 audios from one second to fifteen seconds and is a large-scale dataset of bird sounds collected, containing 10,000/1,400/2,720 in training, validation, and testing datasets, respectively~\cite{zhang2023birdsoundsdenoising}. 

\begin{table}
  \caption{Comparison results on the VoiceBank-DEMAND dataset. ``$-$" means not applicable.}
  \label{tab:voice}
  \centering
  \vspace{-0.2cm}
\setlength{\tabcolsep}{+1mm}{
  \begin{tabular}{lcllllll}
    \toprule
    Methods & Domain & PESQ & STOI & CSIG & CBAK & COVL \\
    \hline
CP-GAN~\cite{liu2020cp} &T &2.64 &0.942 &3.93& 3.33 &3.28\\
PGGAN~\cite{li2022perception} &T &2.81 &0.944& 3.99& 3.59 &3.36\\
DCCRGAN~\cite{huang2022dccrgan} &TF &2.82 &0.949 &4.01& 3.48 &3.40 \\
S-DCCRN~\cite{lv2022s} &TF &2.84 &0.940 &4.03 &2.97 &3.43 \\
DCU-Net~\cite{choi2019phase} &TF &2.93 &0.930 &4.10 &3.77& 3.52 \\
PHASEN~\cite{yin2020phasen} &TF &2.99 &$-$ &4.18& 3.45& 3.50 \\
MetricGAN+~\cite{fu2021metricgan+} &TF &3.15 &0.927& 4.14& 3.12 &3.52 \\
TSTNN~\cite{wang2021tstnn} & T & 2.96 & 0.950 & 4.33 & 3.53 & 3.67 \\
MANNER~\cite{park2022manner} & T & 3.21 & 0.950 & 4.53 & 3.65 & 3.91 \\
\hline
DCHT & T+TF & \textbf{3.41} & \textbf{0.954} & \textbf{4.78} & \textbf{3.82} & \textbf{4.22} \\
\hline
  \end{tabular}}
  \vspace{-.5cm}
\end{table}

\subsection{Implementation details} 
Our model is an end-to-end trainable model without any pre-trained networks and implemented by PyTorch with a single NVIDIA GTX 3060 GPU. All the utterances are resampled to 16 kHz. Each frame has a size of 512 samples (32ms) with an overlap of 256 samples (16ms). Within a batch, the smaller utterances are zero-padded to match the size of the largest utterance. We select the best model using the validation dataset.

At the training stage, we train our model for 100 epochs and optimize it using the Adam algorithm. We use gradient clipping with a maximum of L2-norm of 5 to avoid gradient explosion~\cite{wang2021tstnn}. For learning rate, we use the dynamic strategies during the training stage~\cite{vaswani2017attention}. 

\subsection{Evaluation metrics}\label{sec:me}
We evaluate the proposed speech enhancement model on the VCTK+DEMAND dataset using several objective metrics: PESQ, STOI, CSIG, CBAK, and COVL, for overall quality evaluation. A structure similarity is also used. For evaluation on the Birdsoundsdenoising dataset, we apply signal-to-distortion ratio (SDR)~\cite{zhang2023birdsoundsdenoising} to evaluate our model. 

\begin{table}
\small
\begin{center}
\caption{Results comparisons of different methods ($F1, IoU$, and $Dice$ scores are multiplied by 100. ``$-$" means not applicable. }
\label{tabcom}
\vspace{-0.2cm}
\setlength{\tabcolsep}{+1mm}{
\begin{tabular}{lllll|lllllllll}
\hline \label{tab:md}
 \multirow{2}{*}{Networks}
 &  \multicolumn{4}{c}{Validation} & \multicolumn{4}{c}{Test}\\
 \cmidrule{2-9}
& $F1$ & $IoU$  & $Dice$ & $SDR$ & $F1$ & $IoU$  & $Dice$ & $SDR$\\
\hline
U$^2$-Net~\cite{qin2020u2}  &60.8 &45.2 &60.6 &7.85 & 60.2  &44.8 &59.9 & 7.70\\
MTU-NeT~\cite{wang2022mixed}  &69.1 &56.5 &69.0  &8.17 & 68.3  &55.7 & 68.3 &7.96\\
Segmenter~\cite{strudel2021segmenter}& 72.6  & 59.6 & 72.5 & 9.24 & 70.8 & 57.7 & 70.7 & 8.52\\
U-Net~\cite{ronneberger2015u} &75.7 &64.3 &75.7 & 9.44 &74.4 &62.9 &74.4 & 8.92\\
SegNet~\cite{badrinarayanan2017segnet}  &77.5 &66.9 &77.5 & 9.55&76.1 &65.3 &76.2 & 9.43\\
DVAD~\cite{zhang2023birdsoundsdenoising}& 82.6  & 73.5 & 82.6 & 10.33  & 81.6 & 72.3 & 81.6 & 9.96\\
\textbf{DCHT} & \textbf{-}  & \textbf{-} & \textbf{-} &  \textbf{10.49}  & \textbf{-} & \textbf{-} & \textbf{-} & \textbf{10.43}\\
\hline
\end{tabular}}
\end{center}
\vspace{-0.2cm}
\end{table}

\begin{figure}
	\centering
	\includegraphics[width=0.5\textwidth]{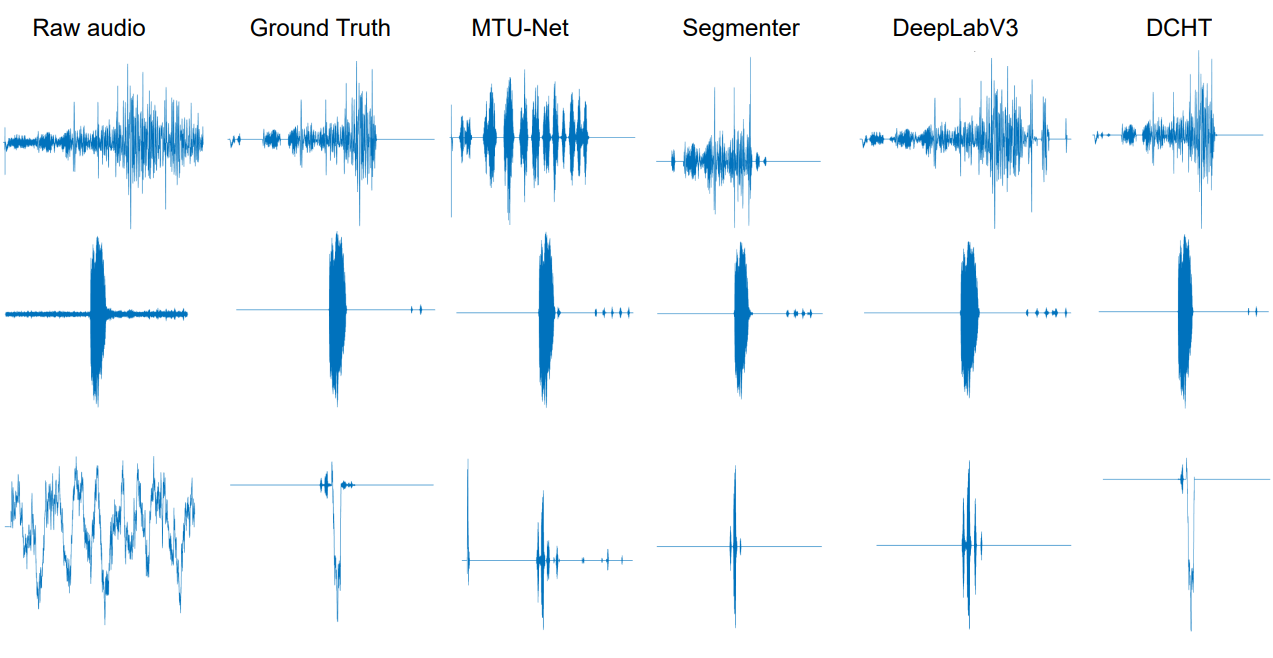}
	\caption{Results comparison on VCTK+DEMAND dataset. Raw audio is the original mixture of audio. Ground Truth is the clean audio.}\label{Fig:audiocom}
\end{figure}

\subsection{Results}
We compare our proposed model with several state-of-the-art baseline models. The results on the BirdSoundsDenoising dataset are shown in Table~\ref{tabcom}~\cite{li2023deeplabv3+}, with the best results in each metric highlighted in boldface. The second to fifth columns are the results of validation, and the latter results are of testing. The results demonstrate that our model outperforms other state-of-the-art methods in terms of SDR. Since these metrics are employed for the audio image segmentation task, the results of F1, IoU, and Dice are ignored~\cite{zhang23p_interspeech}.

We also conduct extra experiments on the VCTK+DEMAND dataset. As shown in Table ~\ref{tab:voice}, the proposed method is compared with other methods in terms of PESQ, STOL, CSIG, CBAK, COVL, and SSIM. The results indicate that our proposed model outperforms other DNNs-based models and achieves state-of-the-art performance in metrics. The comparisons of raw bird audio, ground truth labeled clean audio, and the estimation audio from several models are displayed in Figure~\ref{Fig:audiocom}. Our model produces results more similar to the labeled clean signal.

\subsection{Ablation Studies}
Experimental results in the previous subsection show that our method improves SE performance. To further verify the effectiveness of our proposed model and show the importance of each component of our model, we perform an ablation study by gradually replacing the components in our model. We designed three experiments labeled the DPTnet-only model, the DCSUnet-only model, and the full model. Table~\ref{abl} shows ablation results for the DPTnet-only model and the CSwinUnet-only model, comparing to the full model. Note that the model performance degrades significantly without the CSwin-Unet module.

\begin{table}[t]
\small
\begin{center}
\caption{\textbf{Ablation Results}. }
\label{abl}
\vspace{-0.2cm}
\setlength{\tabcolsep}{+5mm}{
\begin{tabular}{lllll|lllllllll}
\hline \label{tab:me}
 Model&Metric\\
\hline
ComplexSwinUnet-only &60.8\\
Improve DPTNET-only &69.1\\
Full Model & \textbf{72.6}\\
\hline
\end{tabular}}
\end{center}
\end{table}

\section{Conclusion}
In this paper, we propose a deep complex hybrid transformer for speech enhancement. The DCHT model combines two types of methods: the deep complex Swin-Unet transformer and the dual-path transformer neural network, performing in parallel to exploit a complementary set of features for speech enhancement. The deep complex Swin-Unet transformer improves the Swin-Unet networks for complex-valued spectrum modeling. Similarly, the improved DPTnet applies memory-compress attention to reduce memory usage. The proposed DCHT model is evaluated and compared with several current mainstream deep learning-based SE methods using different datasets. The experimental results show that our model can better remove noise in speech and show significant improvements in evaluation metrics.

\bibliographystyle{unsrt}
\bibliography{ref}

\begin{thebibliography}{10}

\bibitem{lu2022conditional}
Yen-Ju Lu, Zhong-Qiu Wang, Shinji Watanabe, Alexander Richard, Cheng Yu, and
  Yu~Tsao.
\newblock Conditional diffusion probabilistic model for speech enhancement.
\newblock In {\em ICASSP 2022-2022 IEEE International Conference on Acoustics,
  Speech and Signal Processing (ICASSP)}, pages 7402--7406. IEEE, 2022.

\bibitem{pascual2017segan}
Santiago Pascual, Antonio Bonafonte, and Joan Serra.
\newblock Segan: Speech enhancement generative adversarial network.
\newblock {\em arXiv preprint arXiv:1703.09452}, 2017.

\bibitem{cao2022cmgan}
Ruizhe Cao, Sherif Abdulatif, and Bin Yang.
\newblock Cmgan: Conformer-based metric gan for speech enhancement.
\newblock {\em arXiv preprint arXiv:2203.15149}, 2022.

\bibitem{jansson2017singing}
A~Jansson, E~Humphrey, N~Montecchio, R~Bittner, A~Kumar, and T~Weyde.
\newblock Singing voice separation with deep u-net convolutional networks.
\newblock In {\em 18th International Society for Music Information Retrieval
  Conference}, pages 23--27, 2017.

\bibitem{choi2019phase}
Hyeong-Seok Choi, Jang-Hyun Kim, Jaesung Huh, Adrian Kim, Jung-Woo Ha, and
  Kyogu Lee.
\newblock Phase-aware speech enhancement with deep complex u-net.
\newblock In {\em International Conference on Learning Representations}, 2019.

\bibitem{hu2020dccrn}
Yanxin Hu, Yun Liu, Shubo Lv, Mengtao Xing, Shimin Zhang, Yihui Fu, Jian Wu,
  Bihong Zhang, and Lei Xie.
\newblock Dccrn: Deep complex convolution recurrent network for phase-aware
  speech enhancement.
\newblock {\em arXiv preprint arXiv:2008.00264}, 2020.

\bibitem{tan2023cst}
Kaijun Tan, Wenyu Mao, Xiaozhou Guo, Huaxiang Lu, Chi Zhang, Zhanzhong Cao, and
  Xingang Wang.
\newblock Cst: Complex sparse transformer for low-snr speech enhancement.
\newblock {\em Sensors}, 23(5):2376, 2023.

\bibitem{zhang2023complex}
Youshan Zhang and Jialu Li.
\newblock Complex image generation swintransformer network for audio denoising.
\newblock In {\em Interspeech}, 2023.

\bibitem{macartney2018improved}
Craig Macartney and Tillman Weyde.
\newblock Improved speech enhancement with the wave-u-net.
\newblock {\em arXiv preprint arXiv:1811.11307}, 2018.

\bibitem{rethage2018wavenet}
Dario Rethage, Jordi Pons, and Xavier Serra.
\newblock A wavenet for speech denoising.
\newblock In {\em 2018 IEEE International Conference on Acoustics, Speech and
  Signal Processing (ICASSP)}, pages 5069--5073. IEEE, 2018.

\bibitem{baby2019sergan}
Deepak Baby and Sarah Verhulst.
\newblock Sergan: Speech enhancement using relativistic generative adversarial
  networks with gradient penalty.
\newblock In {\em ICASSP 2019-2019 IEEE International Conference on Acoustics,
  Speech and Signal Processing (ICASSP)}, pages 106--110. IEEE, 2019.

\bibitem{kong2022speech}
Zhifeng Kong, Wei Ping, Ambrish Dantrey, and Bryan Catanzaro.
\newblock Speech denoising in the waveform domain with self-attention.
\newblock In {\em ICASSP 2022-2022 IEEE International Conference on Acoustics,
  Speech and Signal Processing (ICASSP)}, pages 7867--7871. IEEE, 2022.

\bibitem{yu2022setransformer}
Weiwei Yu, Jian Zhou, HuaBin Wang, and Liang Tao.
\newblock Setransformer: speech enhancement transformer.
\newblock {\em Cognitive Computation}, pages 1--7, 2022.

\bibitem{kim2021kuielab}
Minseok Kim, Woosung Choi, Jaehwa Chung, Daewon Lee, and Soonyoung Jung.
\newblock Kuielab-mdx-net: A two-stream neural network for music demixing.
\newblock {\em arXiv preprint arXiv:2111.12203}, 2021.

\bibitem{defossez2021hybrid}
Alexandre D{\'e}fossez.
\newblock Hybrid spectrogram and waveform source separation.
\newblock {\em arXiv preprint arXiv:2111.03600}, 2021.

\bibitem{rouard2022hybrid}
Simon Rouard, Francisco Massa, and Alexandre D{\'e}fossez.
\newblock Hybrid transformers for music source separation.
\newblock {\em arXiv preprint arXiv:2211.08553}, 2022.

\bibitem{li2020speech}
Andong Li, Minmin Yuan, Chengshi Zheng, and Xiaodong Li.
\newblock Speech enhancement using progressive learning-based convolutional
  recurrent neural network.
\newblock {\em Applied Acoustics}, 166:107347, 2020.

\bibitem{liu2021swin}
Ze~Liu, Yutong Lin, Yue Cao, Han Hu, Yixuan Wei, Zheng Zhang, Stephen Lin, and
  Baining Guo.
\newblock Swin transformer: Hierarchical vision transformer using shifted
  windows.
\newblock In {\em Proceedings of the IEEE/CVF international conference on
  computer vision}, pages 10012--10022, 2021.

\bibitem{wang2021tstnn}
Kai Wang, Bengbeng He, and Wei-Ping Zhu.
\newblock Tstnn: Two-stage transformer based neural network for speech
  enhancement in the time domain.
\newblock In {\em ICASSP 2021-2021 IEEE International Conference on Acoustics,
  Speech and Signal Processing (ICASSP)}, pages 7098--7102. IEEE, 2021.

\bibitem{luo2020dual}
Yi~Luo, Zhuo Chen, and Takuya Yoshioka.
\newblock Dual-path rnn: efficient long sequence modeling for time-domain
  single-channel speech separation.
\newblock In {\em ICASSP 2020-2020 IEEE International Conference on Acoustics,
  Speech and Signal Processing (ICASSP)}, pages 46--50. IEEE, 2020.

\bibitem{braun2021consolidated}
Sebastian Braun and Ivan Tashev.
\newblock A consolidated view of loss functions for supervised deep
  learning-based speech enhancement.
\newblock In {\em 2021 44th International Conference on Telecommunications and
  Signal Processing (TSP)}, pages 72--76. IEEE, 2021.

\bibitem{valentini2016speech}
Cassia Valentini-Botinhao, Xin Wang, Shinji Takaki, and Junichi Yamagishi.
\newblock Speech enhancement for a noise-robust text-to-speech synthesis system
  using deep recurrent neural networks.
\newblock In {\em Interspeech}, volume~8, pages 352--356, 2016.

\bibitem{veaux2013voice}
Christophe Veaux, Junichi Yamagishi, and Simon King.
\newblock The voice bank corpus: Design, collection and data analysis of a
  large regional accent speech database.
\newblock In {\em 2013 international conference oriental COCOSDA held jointly
  with 2013 conference on Asian spoken language research and evaluation
  (O-COCOSDA/CASLRE)}, pages 1--4. IEEE, 2013.

\bibitem{zhang2023birdsoundsdenoising}
Youshan Zhang and Jialu Li.
\newblock Birdsoundsdenoising: Deep visual audio denoising for bird sounds.
\newblock In {\em Proceedings of the IEEE/CVF Winter Conference on Applications
  of Computer Vision}, pages 2248--2257, 2023.

\bibitem{liu2020cp}
Gang Liu, Ke~Gong, Xiaodan Liang, and Zhiguang Chen.
\newblock Cp-gan: Context pyramid generative adversarial network for speech
  enhancement.
\newblock In {\em ICASSP 2020-2020 IEEE International Conference on Acoustics,
  Speech and Signal Processing (ICASSP)}, pages 6624--6628. IEEE, 2020.

\bibitem{li2022perception}
Yihao Li, Meng Sun, and Xiongwei Zhang.
\newblock Perception-guided generative adversarial network for end-to-end
  speech enhancement.
\newblock {\em Applied Soft Computing}, 128:109446, 2022.

\bibitem{huang2022dccrgan}
Huixiang Huang, Renjie Wu, Jingbiao Huang, Jucai Lin, and Jun Yin.
\newblock Dccrgan: Deep complex convolution recurrent generator adversarial
  network for speech enhancement.
\newblock In {\em 2022 International Symposium on Electrical, Electronics and
  Information Engineering (ISEEIE)}, pages 30--35. IEEE, 2022.

\bibitem{lv2022s}
Shubo Lv, Yihui Fu, Mengtao Xing, Jiayao Sun, Lei Xie, Jun Huang, Yannan Wang,
  and Tao Yu.
\newblock S-dccrn: Super wide band dccrn with learnable complex feature for
  speech enhancement.
\newblock In {\em ICASSP 2022-2022 IEEE International Conference on Acoustics,
  Speech and Signal Processing (ICASSP)}, pages 7767--7771. IEEE, 2022.

\bibitem{yin2020phasen}
Dacheng Yin, Chong Luo, Zhiwei Xiong, and Wenjun Zeng.
\newblock Phasen: A phase-and-harmonics-aware speech enhancement network.
\newblock In {\em Proceedings of the AAAI Conference on Artificial
  Intelligence}, volume~34, pages 9458--9465, 2020.

\bibitem{fu2021metricgan+}
Szu-Wei Fu, Cheng Yu, Tsun-An Hsieh, Peter Plantinga, Mirco Ravanelli, Xugang
  Lu, and Yu~Tsao.
\newblock Metricgan+: An improved version of metricgan for speech enhancement.
\newblock {\em arXiv preprint arXiv:2104.03538}, 2021.

\bibitem{park2022manner}
Hyun~Joon Park, Byung~Ha Kang, Wooseok Shin, Jin~Sob Kim, and Sung~Won Han.
\newblock Manner: Multi-view attention network for noise erasure.
\newblock In {\em ICASSP 2022-2022 IEEE International Conference on Acoustics,
  Speech and Signal Processing (ICASSP)}, pages 7842--7846. IEEE, 2022.

\bibitem{vaswani2017attention}
Ashish Vaswani, Noam Shazeer, Niki Parmar, Jakob Uszkoreit, Llion Jones,
  Aidan~N Gomez, {\L}ukasz Kaiser, and Illia Polosukhin.
\newblock Attention is all you need.
\newblock {\em Advances in neural information processing systems}, 30, 2017.

\bibitem{qin2020u2}
Xuebin Qin, Zichen Zhang, Chenyang Huang, Masood Dehghan, Osmar~R Zaiane, and
  Martin Jagersand.
\newblock U2-net: Going deeper with nested u-structure for salient object
  detection.
\newblock {\em Pattern recognition}, 106:107404, 2020.

\bibitem{wang2022mixed}
Hongyi Wang, Shiao Xie, Lanfen Lin, Yutaro Iwamoto, Xian-Hua Han, Yen-Wei Chen,
  and Ruofeng Tong.
\newblock Mixed transformer u-net for medical image segmentation.
\newblock In {\em ICASSP 2022-2022 IEEE International Conference on Acoustics,
  Speech and Signal Processing (ICASSP)}, pages 2390--2394. IEEE, 2022.

\bibitem{strudel2021segmenter}
Robin Strudel, Ricardo Garcia, Ivan Laptev, and Cordelia Schmid.
\newblock Segmenter: Transformer for semantic segmentation.
\newblock In {\em Proceedings of the IEEE/CVF international conference on
  computer vision}, pages 7262--7272, 2021.

\bibitem{ronneberger2015u}
Olaf Ronneberger, Philipp Fischer, and Thomas Brox.
\newblock U-net: Convolutional networks for biomedical image segmentation.
\newblock In {\em Medical Image Computing and Computer-Assisted
  Intervention--MICCAI 2015: 18th International Conference, Munich, Germany,
  October 5-9, 2015, Proceedings, Part III 18}, pages 234--241. Springer, 2015.

\bibitem{badrinarayanan2017segnet}
Vijay Badrinarayanan, Alex Kendall, and Roberto Cipolla.
\newblock Segnet: A deep convolutional encoder-decoder architecture for image
  segmentation.
\newblock {\em IEEE transactions on pattern analysis and machine intelligence},
  39(12):2481--2495, 2017.

\bibitem{li2023deeplabv3+}
Junhui Li, Pu~Wang, and Youshan Zhang.
\newblock Deeplabv3+ vision transformer for visual bird sound denoising.
\newblock {\em IEEE Access}, 2023.

\bibitem{zhang23p_interspeech}
Youshan Zhang and Jialu Li.
\newblock {Complex Image Generation SwinTransformer Network for Audio
  Denoising}.
\newblock In {\em Proc. INTERSPEECH 2023}, pages 186--190, 2023.

\end{thebibliography}

\end{document}